\documentclass[sigconf]{acmart} %\documentclass[manuscript,review]{acmart})
\AtBeginDocument{%
  \providecommand\BibTeX{{%
    \normalfont B\kern-0.5em{\scshape i\kern-0.25em b}\kern-0.8em\TeX}}}

\setcopyright{acmcopyright}
\copyrightyear{2022}
\acmYear{2022}
\acmDOI{10.1145/1122445.1122456}
 
\acmConference[CHI '22]{CHI '22: CHI Conference on Human Factors in Computing Systems}{April 30 -- May 6, 2022}{}
\acmPrice{15.00}
\acmISBN{978-1-4503-XXXX-X/18/06}

\usepackage{graphicx} % ``demo`` option just for this example
\usepackage{subcaption}
\usepackage{makecell}
\usepackage{tabularx}

\newcommand{\edit}{\textcolor{black}}
\newcommand{\blue}{\textcolor{black}}

\usepackage{color-edits}
\addauthor{hf}{blue} % Hao-Fei
\addauthor{ls}{purple} % Logan
\addauthor{vs}{brown} % Venkat
\addauthor{ak}{green} % Anna
\addauthor{hz}{red} % Haiyi
\addauthor{sw}{orange} % Steven
\addauthor{kh}{olive} % Ken
\addauthor{ap}{pink} %Adam
\addauthor{placeholder}{red}

\copyrightyear{2022}
\acmYear{2022}
\setcopyright{rightsretained}
\acmConference[CHI '22]{CHI Conference on Human Factors in Computing Systems}{April 29-May 5, 2022}{New Orleans, LA, USA}
\acmBooktitle{CHI Conference on Human Factors in Computing Systems (CHI '22), April 29-May 5, 2022, New Orleans, LA, USA}
\acmDOI{10.1145/3491102.3517439}
\acmISBN{978-1-4503-9157-3/22/04}
\begin{document}

\title[Improving Human-AI Partnerships in Child Welfare]{Improving Human-AI Partnerships in Child Welfare: Understanding Worker Practices, Challenges, and Desires for Algorithmic Decision Support}

\author{Anna Kawakami}
\affiliation{%
  \institution{Carnegie Mellon University}
  \city{Pittsburgh}
  \country{USA}
}
\email{akawakam@andrew.cmu.edu}

\author{Venkatesh Sivaraman}
\affiliation{%
  \institution{Carnegie Mellon University}
  \city{Pittsburgh}
  \country{USA}
}
\email{vsivaram@andrew.cmu.edu}

\author{Hao-Fei Cheng}
\affiliation{%
  \institution{Carnegie Mellon University}
  \city{Pittsburgh}
  \country{USA}
}
\email{haofeic@andrew.cmu.edu}

\author{Logan Stapleton}
\affiliation{%
  \institution{University of Minnesota}
  \city{Minneapolis}
  \country{USA}
}
\email{stapl158@umn.edu}

\author{Yanghuidi Cheng}
\affiliation{%
  \institution{Carnegie Mellon University}
  \city{Pittsburgh}
  \country{USA}
}
\email{yangcheng@cmu.edu}

\author{Diana Qing}
\affiliation{%
  \institution{University of California, Berkeley}
  \city{Berkeley}
  \country{USA}
}
\email{dianaqing@berkeley.edu}

\author{Adam Perer}
\affiliation{%
  \institution{Carnegie Mellon University}
  \city{Pittsburgh}
  \country{USA}
}
\email{adamperer@cmu.edu}

\author{Zhiwei Steven Wu}
\affiliation{%
  \institution{Carnegie Mellon University}
  \city{Pittsburgh}
  \country{USA}
}
\email{zstevenwu@cmu.edu}

\author{Haiyi Zhu}
\affiliation{%
  \institution{Carnegie Mellon University}
  \city{Pittsburgh}
  \country{USA}
}
\email{haiyiz@cs.cmu.edu}

\author{Kenneth Holstein}
\affiliation{%
  \institution{Carnegie Mellon University}
  \city{Pittsburgh}
  \country{USA}
}
\email{kjholste@cs.cmu.edu}

\renewcommand{\shortauthors}{Anna Kawakami et. al.}

\begin{abstract}
AI-based decision support tools (ADS) are increasingly used to augment human decision-making in high-stakes, social contexts. As public sector agencies begin to adopt ADS, it is critical that we understand workers’ experiences with these systems in practice. In this paper, we present findings from a series of interviews and contextual inquiries at a child welfare agency, to understand how they currently make AI-assisted child maltreatment screening decisions. \edit{Overall, we observe how workers’ reliance upon the ADS is guided by (1) their knowledge of rich, contextual information beyond what the AI model captures, (2) their beliefs about the ADS’s capabilities and limitations relative to their own, (3) organizational pressures and incentives around the use of the ADS, and (4) awareness of misalignments between algorithmic predictions and their own decision-making objectives. Drawing upon these findings, we discuss design implications towards supporting more effective human-AI decision-making.}

\end{abstract}


% %% A "teaser" image appears between the author and affiliation
% %% information and the body of the document, and typically spans the

\maketitle
\section{Introduction}

AI-based decision support tools (ADS) are increasingly used to augment human workers in complex social contexts, including social work, education, healthcare, and criminal justice \cite{De-Arteaga2020, Green2019, Holstein, Levy2021, HoltenMoller2020, Yang2019}. These technologies promise to foster better, more equitable decision outcomes by helping to overcome limitations and biases in human judgments \cite{Chouldechova2018, kahneman2021noise, Levy2021}. At the same time, AI-based judgments are themselves likely to be imperfect and biased, even if in different ways than humans \cite{De-Arteaga2020, Levy2021, veale2018fairness, keddell2019algorithmic}. Given the potential for humans and AI systems to build upon each others’ strengths, and to compensate for each others’ limitations, a growing body of research has sought to design for effective \textit{human--AI partnerships}: configurations of humans and AI systems that can draw upon complementary strengths of each \cite{bansal2021does, De-Arteaga2020, Holstein, Kamar2016, Tan}. Yet to date, little is known about what factors might foster or hinder effective human--AI partnerships in practice, across different real-world contexts.

In this work, we investigate how social workers at a US-based child welfare agency currently make AI-assisted child maltreatment screening decisions in their day-to-day work. In one form or another, AI-based decision supports are anticipated to play an important role in the future of child welfare decision-making \cite{aclu2021family, brown2019toward, Chouldechova2018, nissen2020social, saxena2020human}. Yet the use of ADS in child welfare remains contentious, and it is unclear what forms of human--AI partnership might be most effective and appropriate in this context \cite{brown2019toward, Chouldechova2018, hurley2018can, Jackson2017, Saxena2021}. While some prior work has focused on understanding families' and other affected community members’ perspectives and desiderata for the use of ADS in child welfare \cite{brown2019toward, cheng2021soliciting}, almost no research has investigated \textit{workers’} experiences working with these systems in practice.

We present findings from a series of interviews and contextual inquiries with child maltreatment hotline workers, aimed at understanding their current practices and challenges in working with an ADS day-to-day. \edit{We examine the use of the Allegheny Family Screening Tool (AFST). The AFST was deployed in Allegheny County in 2016, to assist child maltreatment hotline workers in assessing risk and prioritizing among referred cases \cite{AFSTdocumentation,Chouldechova2018}. The AFST context has been frequently studied in recent years (e.g., \cite{cheng2021soliciting, Chouldechova2018, De-Arteaga2020, vaithianathan2021using}), and public sector agencies are beginning to look to the AFST as an example of what AI-assisted decision-making can or should look like in child welfare and similar contexts \cite{aclu2021family}. However, most prior research on the AFST has relied on retrospective quantitative analyses of workers’ decisions, without an understanding of how workers actually integrate the AFST into their decision-making on-the-ground. 
In this work, we focus on understanding how workers currently use the AFST in their day-to-day work, and what design opportunities exist to support more effective AI-assisted decision-making. We explore the following research questions:
\begin{enumerate}
\item How do workers decide when, whether, and how much to rely upon algorithmic recommendations? 
\item What limitations and future design opportunities do workers perceive for the AFST or future ADS tools?
\end{enumerate}   }

\edit{We found that, although the AFST had been in use for half a decade, the system remained a source of tension for many workers,} who perceived the system's current design as a missed opportunity to effectively complement their own abilities. As a step towards the design of new forms of human-AI partnership in child welfare, we engaged these practitioners in envisioning how future technologies might better support their needs. \edit{In the remainder of this paper, we first provide a brief overview of related work and describe the child welfare decision-making context in which this work is situated. We then describe our contextual inquiries, semi-structured interviews, and analysis approach. 
Based on our analysis, we present rich findings capturing workers' current practices and challenges in working with the AFST. 
We discuss how workers’ reliance upon the AFST is guided by (1) their knowledge of rich, contextual information beyond what the AI model captures, (2) their beliefs about the ADS’s capabilities and limitations relative to their own, (3) organizational pressures and incentives that they perceive around the use of the ADS, and (4) workers' awareness of misalignments between the ADS's predictive targets versus their own decision-making objectives.
Based on our findings, we present directions for future research and design implications towards supporting more effective human-AI decision-making in child welfare and beyond. Taken together, this work contributes to ongoing discussions in the literature (e.g., \cite{ammitzboll2021street, HoltenMoller2020, Saxena2021}) regarding the need for a broader re-consideration of how ADS should be designed, evaluated, and integrated into public sector contexts.}

\edit{This work represents the first in-depth qualitative investigation in the literature of workers' current practices and challenges in working with the AFST. Overall, our findings complicate narratives from prior academic and grey literature regarding how the AFST fits into workers' day-to-day decision-making. We expect that the kinds of challenges discussed throughout this paper are not uncommon across AI-assisted public sector decision-making contexts. However, it is uncommon for public sector agencies to open their doors to researchers. We recognize Allegheny County for their strong commitment to transparency, for allowing researchers to closely observe their practices, and for their receptiveness to exploring ways to improve their current practices. We hope that this approach will become a norm in the design, development, and deployment of public sector ADS more broadly.}

\section{Background and Related Work}

\subsection{Designing effective human-AI partnerships}
%Against the backdrop of public discourse on the role of algorithms and AI in high-stakes decision making contexts, 
As AI systems are increasingly used to support human work across a range of high-stakes decision making contexts, a growing body of research has sought to design for effective \textit{human–AI partnerships}: configurations of humans and AI systems that draw upon complementary strengths of each \cite{bansal2021does, De-Arteaga2020, Holstein, Kamar2016, Tan}. To date, scientific and design knowledge remains scarce regarding what factors foster or hinder effective human–AI partnerships in practice, across different real-world contexts. 
%However, a complete picture of how these partnerships affect decision-making across different contexts remains elusive
In some studies, human and machine intelligence combined has been shown to outperform either humans or AI alone \cite{bansal2021does, Holstein2018, Patel2019}, while in others human–AI partnerships have failed to improve or have even harmed decision quality, for various reasons \cite{Green2019,Poursabzi-Sangdeh2021}. For instance, a long line of literature demonstrates that humans are often either \textit{too skeptical} of useful ADS outputs or \textit{too reliant upon} erroneous or harmfully biased AI outputs (e.g., \cite{buccinca2021trust, dietvorst2015algorithm, Green2019, lee2004trust}. Much recent research has utilized large-scale crowd experiments on platforms such as Amazon's Mechanical Turk, to investigate how people integrate algorithmic predictions and recommendations into their decision-making (e.g.,~\cite{bansal2021does,buccinca2021trust,Poursabzi-Sangdeh2021,Tan}). However, experiments with crowdworkers performing simulated tasks may be limited in what they can teach us about decision-making in complex real-world decision contexts such as in social work, criminal justice, education, and healthcare~\cite{LurieMulligan2020,Tan}.
%experiments with largely untrained workers performing simulated tasks may fail to capture the nuances of decision making in real-world contexts.

A complementary line of research in HCI has studied and designed for human–AI partnerships in real-world work settings (e.g.,~\cite{De-Arteaga2020,Holstein2019,HoltenMoller2020,Smith2020,VandenBroek2020,Yang2019}). For example, Yang et al.~\cite{yang2016investigating, Yang2019} investigated why efforts to integrate AI-based decision support tools often fail in clinical settings. Building on their findings, the authors designed a radically new form of decision support that better aligned with healthcare workers' actual needs, routines, and organizational workflows. Similarly, through participatory design workshops with public service (job placement) caseworkers, data scientists, and system developers, Holten Møller et al. \cite{HoltenMoller2020} found that while data scientists were fixated on developing decision support tools that predicted individual's risk of long-term unemployment, workers desired fundamentally different forms of decision support.
As discussed in the next subsection, relatively few research projects have explored how to effectively design human–AI partnerships for social work contexts \cite{De-Arteaga2020,Saxena2021}.
%van den Broek et al.'s study on hiring algorithms \cite{VandenBroek2020} and Yang et al.'s work on medical implant decisions \cite{Yang2019} contributed in-depth ethnographic analyses of decision makers -- as well as those adjacent to those decisions -- to identify challenges for the deployment of ADS for high-stakes decisions.

%MAYBE INSERT INTO NEXT SUBSECTION, WITH EMPHASIS ON DE-ARTEAGA WORK --> Recent field research by PIs Chouldechova and Holstein provides early evidence that, when human workers are empowered to evaluate and (as appropriate) second-guess ADS predictions, this may support more effective and equitable decision-making (De Arteaga et al., 2020; Holstein et al., 2018b; 2019a; 2021). Yet it remains an open research question how ADS can best be designed to foster such human–algorithm synergy.

\subsection{Algorithmic decision support in child welfare}

Over the past two decades, many child welfare agencies have begun incorporating computerized decision support tools into various stages of the child protection decision-making process \cite{brown2019toward,Johnson2004}. The most widely-used tools resemble simple checklists, assisting social workers in assessing the risk of a given case based on a small set of manually-entered factors \cite{saxena2020human,Saxena2021}. Yet recent years have seen the introduction of machine learning-based systems such as the privately-owned Rapid Safety Feedback program, deployed in child welfare agencies across several US states~\cite{Jackson2017}, and the public Allegheny Family Screening Tool (AFST) in Pennsylvania~\cite{AFSTdocumentation}. These AI-based decision support tools (ADS) make use of hundreds of features that are automatically pulled from multi-system administrative data, promising to support more consistent, fairer decision-making, particularly given that social workers are often operating under limited information and time pressure~\cite{Chouldechova2018}. However, critics have warned that the use of ADS in child welfare risks amplifying harmful biases in the child welfare system (e.g., reflecting biases in the administrative data on which they are trained and operate upon)~\cite{Chouldechova2018,eubanks2018automating} or introducing absurd errors into decision processes, which humans would not otherwise make~\cite{alkhatib2021live,De-Arteaga2020,Jackson2017}.

Several recent attempts to deploy ADS in child welfare have failed due to concerns among affected communities or among the social workers tasked with using these systems (e.g.,~\cite{Jackson2017,Saxena2021}). To avoid such failures, researchers have highlighted a need for greater community participation during the design phase for new ADS, to ensure that concerns are identified and addressed early on. For example, Brown et al. conducted participatory workshops to better understand the concerns of affected communities, and to identify strategies for improving communities’ comfort with ADS~\cite{brown2019toward}. Similarly, Cheng et al. elicited subjective notions of \edit{fairness in child welfare decision-making from parents and social workers across the United States---who did not necessarily have familiarity or experience using existing ADS tools---in order to inform the design of future ADS tools that align with these notions \cite{cheng2021soliciting}.}

In contrast to this prior work, the current paper contributes to a nascent body of practitioner-oriented research in HCI that studies the integration of existing ADS in public sector decision-making contexts, with the goal of informing more successful uses of such technologies in the future~\cite{De-Arteaga2021, Levy2021, Saxena2021}. For example, drawing upon an ethnographic case study with child welfare caseworkers, Saxena et al.~\cite{Saxena2021} presented \edit{a theoretically-derived framework (ADMAPS) to inform the design of algorithmic decision-supports in public sector contexts. The current research takes a bottom-up approach, validating aspects of the ADMAPS framework, while also presenting several complementary findings. Given that our work considers a machine learning-based system that incorporates hundreds of input variables, in contrast to the simpler algorithmic systems studied by Saxena et al, the intelligibility and transparency of the ADS are key concerns in our context}. In another study, De-Arteaga et al.~\cite{De-Arteaga2020} conducted retrospective quantitative analyses of child maltreatment hotline workers' decision-making while using the AFST, finding encouraging evidence that in aggregate, social workers are able to employ their discretion~\cite{alkhatib2019street,Lipsky1980} to compensate for erroneous algorithmic risk assessments. We build upon and extend this prior work to \edit{gain an on-the-ground understanding of child maltreatment hotline call workers' current practices and challenges in working with the AFST day-to-day, and to gain insight into how they may be able to compensate for algorithmic limitations in practice.} Towards the design of new forms of ADS and human-AI partnership in child welfare, we engage these practitioners in envisioning how future technologies might better support their needs.

%While Saxena et al.'s work analyzed participants' use of ADS against their ADMAPS framework, we take a bottom-up approach to identify previously unanticipated obstacles. In addition, because our work considers a machine learning-based system that incorporates hundreds of input variables, interpretability is a key concern that we explore by probing users' mental models of the AFST.

\subsection{The Allegheny Family Screening Tool and workplace context}
\label{AFSTContext}

% Adjusted text from grant proposal - Publicly known CYF background + AFST-specific information (1-ish paragraphs) 

% Meta-review asks us to "revise assumptions about the context and clarify important aspects to cater to a global audience." 2AC says: "describe the U.S. child welfare system a bit more to highlight the role social workers’ decisions can have in the future of children."}
\edit{
In the United States, the term ``child welfare system'' refers to a continuum of service: child protection, family preservation, kinship care, foster care placements, and adoption services. %This system includes both public agencies and contracted private providers, and is highly dependent on other service sectors, such as health systems; financial, housing, and benefit programs; and substance abuse treatment services. 
The child welfare system’s primary purpose is to keep children safe and protect them from harm, activities typically carried out through a series of decisions made in the screening and investigation of abuse or neglect allegations. Its secondary purpose is to connect families to services that will improve conditions in their homes, supporting children at risk. 
Given the number of children who are reported for maltreatment relative to the number who enter foster care, significant staffing resources are dedicated to the ``front-end'' of the system: assessments, screenings, and investigations. At the same time, however, most investments have focused on the ``back-end'' of the system, or those children removed and placed in foster care \cite{haskins2020child}. This attention has come at the expense of other operational aspects of the child welfare system, such as training and decision support for child maltreatment hotline call screeners and supervisors, caseworkers, or other front-end workers. Today, child welfare agencies in the US are typically staffed and led by social workers who entered the workforce because they were interested in clinical, direct practice with children and families. They do not view themselves as investigators, nor has their professional training necessarily prepared them for the investigative work with which they are tasked \cite{naccarato2010child}.
}

\begin{figure*}
    \includegraphics[width=\textwidth]{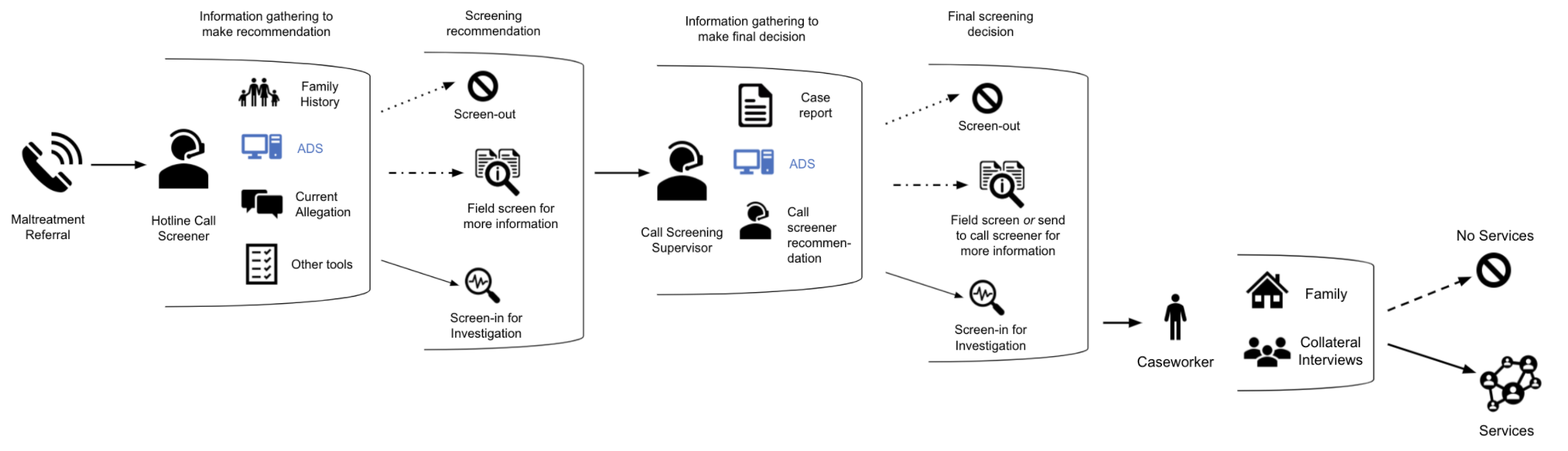}
    \caption{A high-level overview of the child maltreatment screening process at Allegheny County, illustrating when an ADS assists call screeners' and supervisors' screening decisions.}
     \label{fig:screening_overview}
\end{figure*}

Given the volume of child maltreatment referrals, call screeners and their supervisors struggle to make systematic use of the administrative data and case history available to them. The stakes of the decisions these workers make day-to-day cannot be overstated, as workers face a challenging balancing act between ``erring on the side of child safety'' versus ``erring on the side of family preservation'' \cite{Chouldechova2018, scherz2011protecting}. As discussed above, in an effort to augment workers’ abilities to efficiently process and prioritize among cases, child welfare agencies have begun to turn to new ADS tools. Among the most publicized examples, and the ADS in the context of our study, is the \textbf{Allegheny Family Screening Tool (AFST)}. The AFST has been in use at Allegheny County, Pennsylvania's \textbf{Office of Children, Youth and Families (CYF)} Intake/Call Screening Department since 2016, where it assists child protection hotline call screeners in assessing risk and prioritizing among referred cases \cite{AFSTdocumentation,Chouldechova2018}. Figure ~\ref{fig:screening_overview} illustrates the CYF screening and investigation process at Allegheny County. From left to right: an external caller (e.g., a teacher or relative) calls a child maltreatment hotline to make a report, i.e., a \textbf{referral}. A call screener is then tasked with recommending whether or not to screen in the report for investigation. The call screener gathers various sources of information to run the AFST, which outputs a score between 1 (low risk of future placement) to 20 (high risk of future placement). Using the AFST score, the current allegation from the caller, and other information sources from public records, the call screener makes a screening recommendation. The call screener may either agree or disagree with a low or high AFST score, when making their recommendation. The supervisor receives the case, along with a case report including the call screener's recommendation, the AFST score, and other case-related information, to make a final decision. The supervisor may then either agree or disagree with the call screener and/or the AFST score. If the supervisor wishes to screen out a case for which the AFST score is 18 or higher (a \textbf{mandatory screen-in score}), they must go through an \textbf{override process} to make their screening decision. If the case is screened in instead, it is referred to a caseworker, who might proceed in a number of different directions (e.g., further observation, investigation, or intervention). If the supervisor finds that there is not enough information to make a screening decision, they may send the report back to the call screener to gather more information (e.g., by calling the reporting source).

%or they may decide to field screen the case (e.g., \placeholdercomment{insert description of field screen here}. 

% \placeholdercomment{todo: check where low-risk and high-risk protocol plays in and edit text as needed. also: add descriptions of placement and re-referral risk? also: add one-sentence about what the low (0-10) / medium (10-15) / high (15-20) cutoffs are}

 % Child welfare agencies are typically staffed and led by social workers who entered the workforce because they were interested in clinical, direct practice with children and families. They do not view themselves as investigators, nor has their professional training adequately prepared them for the investigative work with which they are tasked.

\section{Methods}
To understand how child maltreatment hotline call screeners and supervisors integrate algorithmic predictions from the AFST into their decision-making, we analyze data from two sources: contextual inquiries and semi-structured interviews with workers. 
%and large-scale administrative data documenting historical algorithm-assisted call screening decisions. 
We visited Allegheny County’s CYF where we conducted a total of approximately 37.5 hours of observations and interviews in order to both understand workers' current decision-making processes using the AFST and to identify design opportunities to support more effective human-algorithm decision-making in the future. 

Over the course of two visits, spread across a two-week period, we conducted contextual inquiries and semi-structured post-interviews with a total of nine call screeners and four supervisors. During the contextual inquiries, we observed call screeners taking calls, compiling case reports, running the AFST, and making screening recommendations. We also observed supervisors reviewing call screeners’ recommendations, case reports, and the AFST scores, and then integrating this information to make their final screening decisions. During each visit, members of our research team conducted observations and interviews with different call screeners and supervisors at the Allegheny CYF office. Each call screener or supervisor was observed and interviewed by one to two researchers, and a different set of call screeners and supervisors was present for each of our visits. Before beginning the observations, we provided the call screener or supervisor an overview of the study purpose and methods and obtained their participation and recording consent. Given that we observed and interviewed call screeners during their working hours, some call screeners or supervisors were unable to complete all study activities. A total of nine call screeners and two supervisors participated in the contextual inquiry \edit{(all participants except S3 and S4)}, and nine call screeners and four supervisors participated in the post-interview. \edit{See Table ~\ref{tab:demographics} for aggregated participant demographics. Note that we provide this information in aggregate form to avoid making individual workers identifiable within their workplace.} %We include our contextual inquiry and semi-structured interview protocols as supplementary materials. 

 \subsection{Contextual inquiries}
 Playing the role of “apprentices” shadowing a “master” to learn a trade \cite{beyer1999contextual}, researchers observed call screeners and supervisors as they performed their jobs, asking follow-up questions as needed . For call screeners, this involved observing how they gather information via incoming calls, gather additional information through administrative databases, fill out reports on individual cases, and make screening recommendations with the AFST, amongst other tasks. For supervisors, this involved observing how they made screening decisions using case reports and the AFST score. When making a screening recommendation or decision, the call screeners and supervisors were asked to think aloud~\cite{van1994think} to help the researcher follow their thought processes. To ensure minimal disruption, researchers did not ask questions when call screeners were on a call. Researchers asked call screeners and supervisors questions in-the-moment, during their decision-making process (see Supplementary Materials). After making a screening decision, the call screeners and supervisors were asked any relevant follow-up questions in a semi-structured interview style. In total, we observed call screeners taking 32 calls, 7 of which led to making a screening recommendation using the risk assessment tool. The two supervisors were observed for 1.5 hours total and made 14 screening decisions with the risk assessment tool during that time. All call screeners and supervisors consented to having researchers take notes on observations and interviews.
 
\subsection{Semi-structured interviews}
Towards the end of our contextual inquiries, we conducted a post-interview to (1) validate our observations, (2) gain further insight into call screeners’ and supervisors’ perceptions and practices around the AFST, and (3) understand design opportunities to improve decision-making with the AFST or similar ADS tools in the future. The first iteration of our post-interview protocol consisted of five sections: (1) participant background (e.g., educational and professional background, years of experience working with the AFST, and experience using other algorithmic decision support tools), (2) clarification and validation of findings from our contextual inquiries, (3) understanding worker perceptions of the AFST, (4) understanding worker beliefs and experiences around fairness and bias in the AFST, and (5) understanding where workers perceive opportunities to augment and improve tools like the AFST. After conducting Interpretation Sessions~\cite{beyer1999contextual} to synthesize findings from our first visit, we iterated on the interview protocol. We added additional questions to probe deeper on three topics that had come up repeatedly during our first visit: understanding how call screeners and supervisors learn about how the AFST works and behaves, understanding how workers communicate with each other around the AFST’s outputs, and understanding worker perceptions of the AFST’s effects on overall screen-in rates. 
See 
Supplementary Materials 
for a full list of interview questions participants were asked during each site visit.

All 13 call screeners and supervisors participated (separately) in the post-interview. 12 call screeners and supervisors consented to being audio-recorded and one consented to having notes taken. The average interview time was approximately 47 minutes for the 12 call screeners and supervisors who were audio-recorded. Any call screeners and supervisors from the first visit day who were also present during the second visit day were asked the new questions from the iterated interview protocol. Three call screeners were asked questions from the original interview protocol only, while six call screeners and four supervisors were asked questions from the iterated interview protocol.

\subsection{Analysis}
After each visit, our research team held Interpretation Sessions~\cite{beyer1999contextual} to collaboratively synthesize findings from our contextual inquiries and post-interviews. Following the two visits, we adopted a thematic analysis approach to analyze approximately 9.5 hours of transcribed audio recordings and 92 pages of notes. Following a shared open coding session to calibrate coding granularity, each participant’s data were qualitatively coded by two to three researchers~\cite{mcdonald2019reliability}. For each participant’s data, we ensured that at least one of the coders was the researcher who conducted the contextual inquiry and/or post-interview with that participant. During this phase, each coder remained open to capturing a broad range of observations in their codes, while also keeping watch for observations related to our original research questions and the six major sections included in the post-interview. 

After resolving any disagreements amongst the coders, we conducted a bottom-up affinity diagramming process~\cite{beyer1999contextual} 
\edit{to iteratively refine and group the resulting 1,529 unique codes into successively higher-level themes. In total, this process generated four levels of themes. The first level clustered our 1,529 codes into 380 themes. These were then clustered into 71 second-level themes, 14 third-level themes, and four fourth-level themes.}

\edit{
The four top-level themes that emerged from this analysis correspond to subsection headers in the Findings section. Each top-level theme captures workers’ existing practices with the AFST, as well as design opportunities that they perceive for the AFST to better support their work. For example, within the fourth-level theme on workers’ beliefs about the AFST (see Section \ref{beliefs}), there are two third-level themes: One focuses on workers’ strategies for gaining greater insight into the AFST's behavior, while the other focuses on characterizing workers’ current beliefs about the AFST and the impacts that these have on their day-to-day practice. The former third-level theme includes 11 second-level themes, for example, capturing workers' specific motivations to learn more about the AFST or describing how they collectively make sense of the AFST model through interactions with other workers. Key findings under each of our four top-level themes are presented in the next section.
}

\subsection{\edit{Ethics and Participant Safety}}
\edit{
We assured all workers that their participation was completely voluntary and that their responses would be kept anonymous. To ensure that workers who participated in the study would not be identifiable within their workplace, we only report participant demographics at an aggregate level and omit participant IDs for a small number of sensitive quotes. We acknowledge that HCI projects framed as participatory research and design sometimes take up many hours of participants' time, and then yield no tangible follow-through for participants \cite{pierre2021getting}. We intend to continue our collaboration with workers and other relevant stakeholders in Allegheny County to ensure that they can benefit from this research.
}

\subsection{\blue{Positionality}}
\blue{We acknowledge that our experiences and positionality shape our perspectives, which guide our research. We are all researchers working in the United States. Our academic backgrounds range across interdisciplinary fields within Computer Science, including HCI and AI. Some of us have prior experiences studying social work contexts or other public-sector decision-making contexts in the United States but not elsewhere in the world. None of us have been investigated by a child welfare agency nor adopted or involved in the foster care system. In addition, none of us have professional experience in child welfare. All authors except two live in Allegheny County; the other two live in Minnesota and California. To conduct this research, we collaborated with Allegheny County's Child, Youth, and Families Department as external researchers. The analysis and writing were conducted independently from the department.} 

% participant demographics table 
\renewcommand{\arraystretch}{2.0}
\begin{table*}[h]
\begin{tabular}{ p{4.3cm} p{12.7cm} }
\hline
\textbf{Demographic information} & \textbf{Participant counts} \\ \hline
Study participation & \parbox[t]{12.7cm}{Contextual inquiry: All participants except for S3 and S4\\ Semi-structured interview: All 13 participants} \\
Years in current position & \parbox[t]{12.7cm}{Call screeners: <1 year (1), 1-3 years (3), 3-5 years (1), >= 5 years (4) \\Supervisors: >= 5 years (4)} \\ 
Have you ever worked in your current position without the AFST? & \parbox[t]{12.7cm}{Call screeners: Yes (4), No (4), Unsure (1) \\Supervisors: Yes (3), No (1)} \\ 
How long have you worked with the AFST? & \parbox[t]{12.7cm}{Call screeners: Since the worker’s employment (5), Since the AFST’s deployment (3), Unsure (1)\\Supervisors: Since the AFST’s deployment (3), Unsure (1)} \\
Do you have prior experience working in other social work positions? & \parbox[t]{12.7cm}{Call screeners: Caseworker experience (5), Other social work-related experience (2), Undisclosed (2)\\Supervisors: Caseworker experience (2), Undisclosed (2)} \\ \hline
\end{tabular}
\caption{\edit{Participants' self-reported demographics aggregated for call screeners and supervisors.}}
\label{tab:demographics}
\end{table*}

\section{Results}
\edit{The AFST has been described as a tool that \textit{``simply augments the human decision whether to investigate a call alleging abuse or neglect’’} \cite{Cherna2018} and \textit{``does not replace clinical judgment but rather provides additional information to assist in the call screening decision making process’’} \cite{alleghenydhs}. Meanwhile, HCI researchers have often expressed concern that the presence of tools like the AFST will lead to overreliance by workers who might place too much trust in these tools’ recommendations (e.g.,~\cite{bansal2021does, buccinca2021trust, De-Arteaga2020, eubanks2018automating, Poursabzi-Sangdeh2021}). \textbf{Our analysis revealed a more complicated picture that is not fully captured by either of these narratives.} }

\edit{In this section, we present our findings across four subsections, each of which corresponds to one of the four top-level themes that emerged through our analysis. We first discuss how workers calibrate their reliance on the AFST by drawing upon rich, contextual information beyond what the AI model is able to capture (Section ~\ref{contextualInfo}). Then, we discuss how workers’ beliefs about the AFST both shape and are shaped by their day-to-day interactions with the tool (Section ~\ref{beliefs}). We next discuss the impacts that organizational pressures and incentive structures may have on workers’ reliance, independent of their trust in the technology itself (Section ~\ref{orgImpacts}). Finally, we discuss how workers adjust their use of the AFST based on their knowledge of misalignments between the AFST’s prediction task and their own decision-making task as human experts (Section ~\ref{valueMisalign}).}

Throughout this section, call screeners are identified with a ``C,'' and supervisors are identified with an ``S.'' For a small number of sensitive quotes, participant IDs are omitted to provide an additional layer of anonymity.

\subsection{\edit{How workers calibrate their reliance on the AFST based on knowledge beyond the model}} \label{contextualInfo}
\edit{
In our contextual inquiries and interviews, we observed that workers calibrated their reliance on algorithmic recommendations by \textbf{drawing upon their own knowledge of a given referred case, which often complemented the information captured by the AFST model.} Much prior research on human-AI decision-making has focused on settings where the AI system has access to a superset of the task-relevant information available to the human \cite{bansal2021does,De-Arteaga2020,LurieMulligan2020}. Yet in many real-world decision-making settings, humans and AI systems may have access to \textit{complementary} sources of information, opening potential for each to help overcome the other's blindspots \cite{alkhatib2019street,Holstein,kleinberg2018human}. Our qualitative findings validate hypotheses posed in earlier work by De-Arteaga et al. (2020). Through retrospective quantitative analyses of workers' decision-making with the AFST, these authors found that workers were able to reliably detect and override erroneous AFST recommendations. To explain how workers were able to do this, the authors speculated that the workers may have been cross-checking algorithmic outputs against other relevant information that was available to them but not to the AFST model \cite{De-Arteaga2020}. Our observations lend credence to this hypothesis:
we found that to inform their screening decisions, workers paid close attention to qualitative details which were reflected neither through administrative data nor through manual inputs that workers are able to provide to the AFST. }

%Note from Ken: Haiyi had noticed that it was unclear why we say "causal" narratives below. Taking a quick pass to try clarifying this (starting at the second sentence of the below paragraph).

\edit{\textbf{Based in part on phone conversations with individuals connected to a case, call screeners constructed rich, causal narratives around a given case, accounting for contextual factors that the AFST overlooks} such as potential motives of the callers who are filing a report or cultural misunderstandings that may be at play. In turn, call screeners often communicated such inferences to supervisors, to support their interpretation and decision-making. 
%new text starts here
Observations during our contextual inquiries revealed that, when triangulating across multiple sources of information, call screeners frequently reasoned about possible \textit{causes} behind the evidence they were seeing. For example, as discussed below, workers reasoned about whether a string of re-referrals for a given case (which can increase the AFST risk score \cite{AFSTdocumentation}) may have been caused by parents repeatedly reporting one another in the midst of a dispute, rather than indicating actual risk to a child.
Similarly, workers did not only consider the statistical risk that a factor %(such as a parent's criminal history) 
might pose to a child in theory, as the AFST does, but also reasoned about the plausibility that this risk would be realized in a specific case. For example, in one instance we observed, a call screener saw that a parent had criminal history in the past but a clean record over the past decade. In the context of other indications that the parent had ``started a new life,'' the call screener judged that the parent's prior criminal history was not relevant to assessing risk. In addition, if a worker knew that a given parent did not have any contact with a child, they would take this information into account, disregarding potential risk factors tied to that parent.
These observations align with recent theoretical frameworks that characterize the role of human discretion in algorithm-assisted public-sector decision-making \cite{alkhatib2019street, Saxena2021} and describe gaps between human and AI-based decision-making \cite{lake2017building}: call screeners were able to reason about potential causal factors based on their on-the-ground knowledge and expertise, complementing the AFST's use of large-scale statistical patterns.
%new text ends here
}

\edit{
\textbf{Both call screeners and supervisors asserted that they decide whether or not to override the AFST primarily based on qualitative details of the allegations, and their understanding of the social context in which those allegations are being made}. As S1 put it, the decision of whether to screen-in or override a high AFST score depends entirely on \textit{``...what's being reported and what the allegations are. You know, and [are] there any safety [concerns] in reference to the younger children in the house?’’} (S1). Another supervisor, S2, observed that the AFST is prone to giving high scores for cases that have been frequently referred, and emphasized the importance of considering the allegation information alongside the AFST score, given that the AFST \textit{``[doesn’t consider] the whole story''}. 
Workers shared specific prior experiences in which a high AFST score appeared to reflect a failure to account for relevant context. For example, C7 shared a case in which the AFST score seemed to be high mainly due to the erroneous inclusion of a child’s father in administrative inputs to the AFST model. In this case, the call screener was aware that the father was in jail, and in fact had no contact with the child. Based on this knowledge, and an absence of other evidence of immediate safety concerns, workers chose to override the AFST. In a different case, the AFST provided a low risk score, but workers decided to screen-in the case based on concerning details in the caller’s allegations, pertaining to a relative who was visiting with the child. In this case, the AFST was not aware that the relative was visiting, and was unable to factor in the potential safety risk that the relative posed to the child.
}

\edit{
We observed that workers also frequently took into account the social context in which a set of allegations were being made. For instance, when making screening decisions, workers reasoned about the relationship of the caller making an allegation to the child, the family, and the alleged perpetrators, as well as the caller’s cultural background and potential motivations. During our contextual inquiries, a few call screeners encountered cases that they believed had high AFST scores mainly because these cases had been re-referred several times, not because the child was truly in any danger. Rather, based on their phone conversations with individuals connected to a case, these call screeners believed that these multiple re-referred cases represented ``retaliation reports’’ (e.g., parents repeatedly reporting one another in the midst of a dispute) or cross-cultural misunderstandings on the part of the caller (see Section ~\ref{beliefImpacts}). 
}

\edit{
\textbf{In contrast to workers’ focus on details of the allegations, the AFST is only able to account for the current allegations at a coarse-grained level.} The AFST factors in a categorical ``allegation type’’ variable, taking into account the broad types of allegations made in both current and past referrals. For example, workers can select whether the allegation in a given case maps to categories such as “Child Behaviors,” “Caregiver Substance Abuse,” ``No/Inadequate Home,’’ ``Neglect,’’ ``Physical Altercation,’’ or ``Parent/Child Conflict’’ 
%“Imminent Risks,” ``Physical Maltreatment,’’ ``Sexual Abuse or Exploitation,’’ ``Sexual Contact Between Children,’’ ``Truancy,’’ or ``Unwilling or Unable to Provide Care’’ 
\cite{AFSTdocumentation}. Workers perceived allegations related to ``Imminent Risks’’ or ``Caregiver Substance Abuse’’ as posing immediate safety risks to the child, and therefore strong indicators of potential child maltreatment; on the other hand, they viewed allegations related to ``Parent/Child Conflict’’ as requiring more context, given that not all forms of conflict between parents and their children are necessarily cause for alarm, and cross-cultural differences may influence how conflicts are perceived. 
}

\edit{
The inclusion of categorical variables that capture aspects of the allegations was added to the AFST in 2018, in response to workers’ desires to have allegation information factor into the AFST’s risk score calculation \cite{AFSTdocumentation}. \textbf{However, given a fixed set of allegation types available for workers to select, workers are currently limited in their ability to communicate relevant contextual knowledge to the model via this mechanism.} As such, the inclusion of ``allegation type’’ variables does not eliminate the need for workers to calibrate their reliance on the model based upon their personal knowledge of a case. Despite these limitations, it is worth noting that allegation type is currently one of the few places in a referral where call screeners have discretion to influence the risk assessment score based on their own interpretations. Workers desired the ability to better incorporate their knowledge of relevant context into the score calculation. For example, C3 proposed that it would be helpful if they could leverage their knowledge of case-specific context to explore counterfactuals, by \textit{``[removing] some things out of the score [... to] kinda play with it and say, `if you take this out of there, what kind of score would this person get?’ ’’} 
}

\edit{
 We next discuss how workers’ perceptions of the AFST’s capabilities and limitations, including its inability to account for many qualitative details of individual referrals, impact how they work with the AFST. 
}

\subsection{\edit{How workers’ beliefs about the AFST shape and are shaped by their day-to-day use of the algorithm, in the absence of detailed training}} \label{beliefs}

\edit{
Given limited formal opportunities to learn how the AFST works, we found that social workers improvised ways to learn about the AFST’s capabilities and limitations themselves---both individually, through their day-to-day use of the tool, and collectively, by sharing their observations and inferences about the tool between workers. In turn, the beliefs that workers develop about the AFST play a major role in shaping how they work with (and around) the algorithm. Below, we first describe how workers attempt to learn about the AFST’s capabilities and limitations, and discuss their desires for opportunities to learn more about how the AFST works. We then present aspects of workers’ mental models of the AFST---i.e., the internal representations or systems of beliefs that they build up through experience, which they use to understand, explain, and predict algorithmic behavior, and then act accordingly \cite{kulesza2012tell,Bansal2019,koch2018group,van2011team}---and we discuss the impacts that these beliefs have on their decision-making and communicative practices around the AFST.
}

\subsubsection{\edit{How workers (try to) learn about the AFST’s capabilities and limitations through day-to-day use of the tool}} \label{beliefFormation} 
\edit{\textbf{\\ Despite having used the AFST day-to-day for multiple years, we found that most call screeners and supervisors know very little about how the AFST works or what data it relies upon.}} 
In one instance, a call screener (C3) turned to the interviewer to ask if they knew how the AFST worked, hoping that they could answer their longstanding questions about what features the AFST uses. \edit{Workers described receiving surface-level procedural training on how to run and view the AFST score a number of years prior (S1, S2, S3, C3, C8, C2), but few could recall being presented with authoritative information on details such as the predictive features used by the AFST, or how these features were weighted.} Moreover, workers’ understanding is not aided by the AFST interface, \edit{which simply provides a numerical risk score or a mandatory screen-in/out message.} 

\edit{
While Allegheny County is outwardly transparent about the AFST’s design and development process, with public-facing materials and documentation published on the web (e.g., \cite{afstfaq,AFSTdocumentation}), \textbf{workers are intentionally provided with minimal information about the AFST model to discourage `gaming the system’ behavior}. That is, the designers and administrators of the AFST had worried that if too much transparency were provided to workers, this might enable workers to strategically manipulate data inputs used by the algorithm, in order to get a desired output. While this is a real concern in algorithm-assisted child welfare decision-making contexts \cite{Saxena2021}, we observed that \textbf{given minimal information about the AFST’s capabilities, limitations, and overall functioning, workers took matters into their own hands, improvising ways to learn more about the tool themselves.} 
%workers provided accurate administrative information but improvised other ways to learn more about the tool themselves. 
}

\edit{
Through their day-to-day interactions with the AFST, workers built up intuitions about the AFST’s behavior and hypotheses about its limitations and biases. For example, \textbf{call screeners described a collaborative, game-like approach to predicting AFST scores}, recalling past occasions where call screeners would informally \textit{``gamble on what this score's gonna be’’} (C7).
\begin{quote}
\textit{``What we do is we have to research everything and… start up a report. And then I see, `Oh, wow. They have a lot of history. Oh, they've got this going on and they have that going on... ’ So then I'll say, I bet this score's gonna be…’’} (C3)
\end{quote}
}

\edit{
In addition to honing their ability to predict AFST scores through these guessing games, \textbf{workers would sometimes discuss the AFST score for a given case amongst themselves, in order to collaboratively make sense of it.} Workers did so informally, even though the screening protocol officially assigns only one call screener and one supervisor per case. Describing their open workspace and collaborative decision-making process, C8 said \textit{``I hear everything [...] we’re sitting on each other’s laps’’}. Similarly, S1 explained that \textit{``it’s not like we all work in a little bubble [...] If we’re struggling with something, we will [...] have a little, brief discussion [...] We say, `Well hey, this score is saying it's a 19, but we don't really see anything on this page based on these allegations that says it should be a 19.’’’}
}

\edit{
Workers also developed strategies to gain a more direct window into the AFST’s behavior. For example, in order to learn more about the impacts that particular factors have on the AFST score, both call screeners and supervisors described \textbf{computing scores on slightly different versions of the same referral data and then comparing them to draw inferences.}} Given that the AFST reads from saved referral records, workers cannot significantly modify its inputs without sacrificing thoroughness or accuracy of documentation. However, they can still gain some insight into the \edit{impacts that particular factors may have on the score by making small adjustments.} For example, \edit{to understand the impact that a particular family member’s administrative records have on the AFST score, a worker might omit that family member from the AFST score calculation, run the AFST to see the score, and then compute another score with the family member included and compare the two scores. }

%VS: adding this subheading to help skimming
%KH: Thanks, this looks great! I tried adding a second part to the subsection header capture workers' desires for support in learning more about the tool
\subsubsection{\edit{Workers are frequently surprised and confused by the AFST's behavior, and want support in learning about the tool}} \label{beliefsLearn}
\edit{
Echoing prior findings from Eubanks \cite{eubanks2018automating}, these experiences using and tinkering with the tool 
%gave workers confidence in making
enabled workers to make
predictions of the AFST score that were often accurate. Yet importantly, the intuitions that workers had developed around the AFST were imperfect: call screeners and supervisors \textbf{often encountered AFST scores that surprised them, reminding them that their understanding of the tool’s behavior is limited.} In reality, the AFST model makes predictions based on a wide array of factors, and C6 lamented that call screeners can \textit{``constantly get into a place where we're assuming that [one particular factor] is what has to be driving the score.’’} Workers shared specific instances in which their assumptions were violated. For example,} referencing a belief that the presence of many historical referrals and service engagement leads to higher AFST scores \edit{(see Section ~\ref{beliefImpacts})}, C2 acknowledged that \textit{``just because you have a long history doesn't necessarily mean that you're going to get a big score. So we just guess that that's potentially what it could be. I'm sure there are other factors… but we don't know for sure.’’} In other cases, workers \edit{found that their assumptions about the model were violated} when they observed the AFST score shifting in response to changes in data fields that they had not previously thought the AFST took into account (C7, S3). \edit{For example, during our observation, C7 stated that the allegations have no impact on the AFST score. However, when rerunning the AFST for an existing case with a previous ``Truancy’’ allegation and newly added current allegations on ``Caregiver Substance Abuse’’ and ``Neglect,’’ the call screener observed the score increasing.} 

\edit{While workers sometimes used unexpected scores to adjust their understanding of the model, at other times, they were simply confused when they could not identify potential causes for unexpected behaviors. For example,} one supervisor (S2) noticed the score changing seemingly at random when they ran it multiple times, even from a 14 (moderate risk) to 7 (low risk), with no apparent changes to the algorithm’s inputs. Regardless of what factors led to unexpected model behavior, \textbf{these experiences led some workers to perceive the AFST as hyper-sensitive or unreliable, contributing to their \edit{overall distrust of the tool}} \edit{(C1, C7, S2). C7 compared the expectation that workers should use the AFST, despite lack of insight into how it works, to blind faith: \textit{``not knowing what data's going into it [...] it's more of having, like, religious faith, I guess, you know?’’} }

\edit{
\textbf{Given these challenges, workers desired more opportunities to learn how the AFST works, to empower them to work with the tool more effectively, and to help them avoid making faulty assumptions about AFST scores in particular instances.} As discussed in Section ~\ref{OrgFeedback}, while the AFST includes a field in which workers can submit feedback and questions to the AFST maintainers, workers expressed that they have rarely received responses that they personally found useful and understandable. Several workers believed that having greater insight into how the model works would enable them to better judge when they can trust the score, and ultimately allow them to make better use of the AFST (C3, C6, C7, C9, S3). As C3 put it,} \textit{``If we knew more about how we got to the score, I think I'd pay more attention to how the score is going to help me''}. 
\edit{
One call screener (C6) expressed that they saw greater transparency around the AFST score calculation as an opportunity to balance out power imbalances:} \textit{``just knowing these are the top three things influencing the score or, you know, something to that effect. Yeah, I mean, it feels like there are ways that it could be better integrated into the workflow [to] make it feel collaborative versus telling me what to do’’} (C6). 

At the time of this study, workers were interested in learning more about how the AFST produces its risk assessment score at nearly every stage of the model’s pipeline. For instance, although workers were aware that the AFST relies upon administrative data that they can view through their web interface, \edit{they wanted to know whether it also draws from other county databases, accounting for data that they themselves cannot easily access (C2, C8, S3). Many workers also desired more opportunities to learn about} the specific factors that the AFST is using, and how those factors are weighted \edit{(C2, C3, C6, C7, C8, S1, S2, S3).} For example, workers were interested in learning which individuals in the referral contributed to the score and by how much \edit{(C3, C6, C8, S3), even wondering whether it was possible for deceased relatives to affect the AFST’s assessment of risk to a child (C8). Beyond new forms of training, several workers expressed desires for new decision-time interfaces that could support their current, informal practices for learning about the AFST’s behavior. For example, an augmented version of the AFST interface might explicitly support workers in rapidly exploring multiple counterfactual inputs to the AFST (cf. \cite{wexler2019if}) in order to learn how particular factors impact the AFST score in specific cases (C1, C6, C8, S1).} 

\subsubsection{\edit{Workers' beliefs about the factors used by AFST influence their reliance on the score}} \label{beliefImpacts}
\edit{Workers bring a range of knowledge and beliefs with them to the task of interpreting an AFST score and integrating it into their decision-making, including both their knowledge about the broader context of a case (Section ~\ref{contextualInfo}) and their beliefs about how the AFST score is computed (Section ~\ref{beliefFormation}). In our contextual inquiries and interviews, most workers referenced} beliefs about how the AFST considered the following four factors, among others: the number of re-referrals on a case, the extent of a family’s prior involvement with public services, the size of the family, and the age of the alleged victim(s). \edit{During our observations, \textbf{workers often made guesses about how each of these features might be influencing the AFST score calculation in a given case}, and they reasoned about whether or not a feature’s potential influence on the score was appropriate. \textbf{These inferences, in turn, informed their perceptions of whether the score was ``too high’’ or ``too low’’ for a given case:} \textit{``the more you use [the AFST], you kind of pick up why it will go a certain way and you can kind of use that in addition to what you do know to make an appropriate determination’’} (C1).}

Many workers believed that a case with \textbf{high numbers of re-referrals} consistently resulted in higher scores (C2, C4, C7, C9, S2). \edit{This aligns with the AFST’s use of past referrals as a positively weighted predictive feature \cite{AFSTdocumentation}.} To illustrate, C2 described prior experiences with ``retaliation reports’’ in which parents in custody battles call on each other repeatedly, generating enough referrals that \textit{``eventually, the score is going to be `Accept [for investigation]’ regardless of what [the referral] is.’’} 
\edit{Workers often used their knowledge of the allegation details to override high AFST scores that appeared to be driven by high numbers of re-referrals (see Section ~\ref{contextualInfo}).} 
While some workers understood that the \textbf{``type of allegation’’} factored into the AFST score (C1, S3), others were uncertain \edit{or believed that it did not have much of an impact (C9, S1, S2). In reality, as discussed in Section ~\ref{contextualInfo}}, since 2018, the AFST model has factored in \edit{some information about the allegation through a categorical variable that captures broad categories of allegations for current and past referrals. Workers’ beliefs about the impact of allegation information on the AFST score appeared to influence their use of the tool. For example,} one participant, who believed that allegation type impacted the AFST score, mentioned not wanting \textit{``to go super crazy on allegations’’} \edit{when entering data on the allegation type,} if they thought it would unfairly drive up the AFST score.

\edit{
Workers also believed that the greater a \textbf{family’s degree of involvement with the ``system’’ and ``services’’}---including public mental health and other medical treatment, criminal history, and welfare records---the greater the AFST score would be. In reality, the AFST model does use behavioral health records (with a small positive weight for substance abuse, but a small negative weight for neurotic disorders) as well as criminal history (with positive weights for features related to whether the victim is in juvenile probation currently or in the past year). However, the AFST no longer uses public welfare records as of November 2018 \cite{AFSTdocumentation}. 
When reviewing individual cases, workers often justified their disagreement with the AFST score with reference to their beliefs about the AFST’S use of system involvement as part of its score calculation. For instance,} C9 explained a high AFST score for a family with no referral history by pointing out the father’s criminal record and mother’s drug use. Describing the limitations with quantitatively assessing system involvement as a predictor for child maltreatment, S1 said, \textit{``If I'm someone who accesses the system, meaning... I get public assistance, I've committed a crime... all that stuff stays in the system, and that's the driving factor for that [AFST] score, but that doesn't mean that the subject child in that report is being maltreated or being in danger.''}

Similarly, workers believed that \textbf{reports with more people in them} received higher AFST scores. Workers noticed the impact that the number of people had on the AFST score, both within and across individual case reports. For example, C5 said, \textit{``the more people that are involved with these families, no matter what it's for, the higher their score's gonna be.''} Similarly, C2 described that they often see a case’s AFST's score increase when they add additional people to the report: \textit{``We had one report where the mom... never had CYF history ever before. Dad was added to the report: [the AFST's recommendation] was automatic screen-in because he has kids with other people who have a long history with us.’’} 
\edit{The belief that reports with more people yield broader AFST scores broadly aligns with the AFST model’s use of the number of children, perpetrators, parents, and victims as predictive features \cite{AFSTdocumentation}. However, although the numbers of perpetrators, parents, and victims have positive feature weights, the number of children (of all age groups, from 3 to 18) actually has a negative feature weight. }

\edit{
Given that the AFST does not account for the type of relationship the perpetrator or parent has with the victim, workers expressed concern that when more individuals (along with their respective histories of system involvement) are added to the report, the score may shift in ways that do not reflect actual changes to the child’s actual safety risk. C8 described that workers sometimes have enough contextual information about the individual’s relationship to the victim to assess whether increases in the AFST score are justified. Other times, however, they may not know enough and are left wondering whether the child is truly more at risk because of the added individual: 
\begin{quote}
\textit{``I feel like everybody's got a number, all right? [...] The problem is [all the individuals added to the report are] hooked with [the victim…] And I wonder sometimes to what degree do people who aren't actually living in the house and affecting the children affect the score. [...] the thing is when you're running the score here, we don't know that, you know, Dad hasn't seen little Deon since he was 3 and he's now 15. [...But] some of it you would know’’} (C8). 
\end{quote}
Similarly,} when explaining why they disagreed with the AFST score’s use of number of people as a predictive feature, C3 related the potential impact to her personal life:  \textit{``I have a niece who has three children and three fathers and she's really a good mother. And if a call came in on her, I'm sure it's gonna be a high score [...] that's the part I think is unfair.''} 

Finally, workers believed that \textbf{younger alleged victims} received higher AFST scores. 
\edit{This is largely correct, as the AFST model weights infant, toddler, and preschool victims more heavily than teenage victims \cite{AFSTdocumentation}.} 
Workers were also aware of a protocol in which the AFST automatically screens in high-risk referrals involving children under the age of three. \edit{Unlike the previous factors, this one tended to align with workers' own assessments: in general, workers agreed that younger victims were at greater risk than older ones (C1, C3, S2). However, workers also perceived that \textbf{older alleged victims}, especially teenage victims, were sometimes given unjustifiably high or low AFST scores, depending on if the person had a history of system involvement. As a result, workers pay close attention to the reasons for referral associated with the case, alongside the alleged victim’s history of system involvement, in order to inform or justify decisions to disagree with the AFST: 
\begin{quote}
\textit{``[It is easy to identify] when the score will appear wrong [...] when we have reports of kids that are truant. When we get a high-risk score for a kid, the only reason why the report is being made is because the kid is truant, and truancy doesn't put a 15 or 16-year-old kid at high risk. And, you know, it's based on them pulling information from this program, this program, this program, and that program, [which] does not impact [whether] the kid is going to school or not.’’} (S1). 
\end{quote}
}

\edit{Beyond the features described above, some workers expressed confusion at how the model factors in other features.} For example, workers perceived that the AFST tended to assign higher risk scores to families from underprivileged racial identities and socioeconomic backgrounds (C1, C2, S2). However, they were unsure whether to attribute high AFST scores to race and socioeconomic status directly, or to other correlated features (C3, C7, C9, S1). \edit{For example, while discussing the compounding effects of race and income, S1 said that if the AFST gives a high score: \textit{``[I speculate it must be from disparity because if] I'm not getting money from [public assistance] the computer can't judge me.’’} (S1).} 

\edit{
\textbf{Overall, workers’ intuitions about the model 
%, based on the four primary factors discussed above---the number of re-referrals on a case, the extent of a family’s prior involvement with public services, the size of the family, and the age of the alleged victim(s)---
reflected predictive features that are actually used by the AFST model.} However, given minimal transparency around the model, workers sometimes \textbf{inferred overly general patterns based on their observations} of the AFST's behavior, or \textbf{misjudged the direction of a feature’s influence}. Furthermore, the accuracy of workers' beliefs about the magnitude of particular features' influence remains unclear.}

\subsection{Beyond ``trust'' in AI: Influences of organizational pressures and incentives} \label{orgImpacts}
Prior work has often understood human reliance on algorithmic decision supports in terms of patterns of ``over-reliance,’’ ``under-reliance,’’ or ``appropriate reliance’’, that are shaped according to how well-calibrated an individual’s trust is in a given algorithmic system~\cite{bansal2021does, buccinca2021trust, lee2004trust, Poursabzi-Sangdeh2021}. However, this conceptualization overlooks the potential impacts that organizational pressures and incentive structures may have on workers’ reliance, independent of their trust in the technology itself. 
\edit{In this section, we first describe the impacts of perceived organizational pressures upon workers’ reliance on the AFST: in many cases, workers agreed with the AFST score not because they saw value in the tool or because they trusted the technology, but rather because they perceived organizational pressures to do so. We then discuss the influences of organizational incentives on workers’ motivations to disagree with algorithmic recommendations or to provide feedback towards improving the AFST model.}

\subsubsection{\edit{Organizational performance measures shape workers’ reliance on algorithmic recommendations}} \label{OrgMeasures}
Several call screeners and supervisors perceived \textbf{organizational pressures to avoid disagreeing with the AFST score ``too often.’’} Workers had either heard about or attended monthly meetings where an internal analytics team discusses how often the workers override mandatory screen-in protocols. Workers shared that while they do not know for sure whether the analytics team has a specific ``acceptable'' rate of overrides, they perceived that in some months, they had crossed an unspoken line by overriding too often. \edit{As one supervisor said, \textit{``someone’s watching to make sure that the majority of those high risk protocols are being assigned.’’}
\textbf{Given these perceived pressures, some call screeners and supervisors said that they will sometimes agree with the AFST, going against their own best judgment:} }
\begin{quote}
\textit{``It's not uncommon to be faced with a higher score, but the allegation that is contained in the report is, like, a low risk kind of allegation, you know? [...] The score is high risk protocol because of all of the other stuff that the tool runs, the algorithm gets you a high score. [...But] there's no safety concerns at this point. So then the screener will say, do I have to assign this? Because it says high risk protocol. Do I have to? So I'll help walk through it [with them] because you don't want to be accused of not using the tool, because there is regular oversight from [the] analytics team.''}
\end{quote}
\edit{\textbf{Even when workers did not feel personally affected by these pressures or incentives, they took them into account when reasoning about others’ motivations.}} For example, when reflecting on what would motivate supervisors to disagree with call screeners' recommendations and align with the AFST score instead, a call screener guessed that it may be to \textit{``align with directives from administration.’’}

\edit{Although some workers felt pressure to make decisions against their own judgement, they expressed awareness and concern regarding the impacts their decisions could have on both families and caseworkers.} Workers believed that, if they were to always agree with the AFST, they would have many more screen-in decisions than necessary. When describing the extent to which the AFST may increase screen-in rates, C5 explained \textit{``I think that if we [only followed the AFST], we'd have to hire at least 50\% more [caseworkers] than we have now.''} This led some workers to worry about the impacts increased screen-in decisions might have on caseworkers, who must follow up with an investigation on all cases that are screened in. For example, during our observations of S2, this supervisor decided to screen out a case with a middle-range AFST score of 13, reasoning that they should avoid burdening caseworkers. S2 explained that caseworkers have sometimes complained to them about their decisions, in cases where they screened in high risk protocol cases due to organizational pressures, against their own judgment. They expressed that the situation is \textit{``tricky’’} because they cannot easily explain their decision to the caseworker, given that caseworkers do not see the AFST score. \edit{\textbf{Caught between conflicting pressures from the administration and from caseworkers, some workers wished the AFST tool would provide not just a number, but support in justifying their decision to those whom it would affect} (C6, C7, C9, S2).}

\subsubsection{\edit{``The input does not feel like a two-way street:’’ Workers' motivations to disagree with or provide feedback on algorithmic recommendations are shaped by organizational incentives}} \label{OrgFeedback}
Workers’ decisions to agree with a high-risk protocol AFST score, even in cases where they disagree, may sometimes be \edit{influenced by \textbf{a desire to avoid additional work that they perceive as unnecessary.}} \edit{When deciding to override a high-risk protocol, workers are required to write open-text responses describing their rationale behind the override---a step that the AFST’s designers intended to produce friction and promote worker reflection \cite{AFSTdocumentation}. However, workers did not see value in this override process, which they found tedious.} 

\edit{Moreover, although the open text field was partly designed to give workers an opportunity to provide feedback on potentially erroneous AFST scores, \textbf{workers were unsure whether or how their feedback would actually be used}. For instance, one supervisor expressed discontent that they usually did not receive any acknowledgement that their feedback had been read and considered, contributing to their feeling that writing these ``narratives’’ was a waste of their time, which unnecessarily increased their workload.} Another participant simply said \textit{``there’s no point’’} in providing feedback on the AFST. 

\edit{When workers did receive responses to feedback, \textbf{they perceived that the administration tended to discount feedback regarding potential limitations and improvements to the AFST.} This left workers feeling that their own expertise, as human decision makers, was being undermined while the effectiveness of the AFST was being defended.} Both call screeners and supervisors said that they had received responses that they perceived as unhelpful, dismissive, or even hostile, leading some to \textit{``give up’’} on giving feedback: 
\begin{quote}
\textit{``The input does not feel like a two-way street. It's, like, we are told why we're wrong. And we just don't actually understand algorithms versus, like, maybe you [the workers] are observing something that [our algorithm] missed, you know? [...] in the past, it's felt like when we've taken advantage of that function [to give feedback on the AFST score], it has kind of resulted in just explaining to you why the score is actually right. [...] I think it's kind of been a point of conflict in the past.''}
\end{quote}

\edit{
Given these experiences, workers shared that they have critical discussions about the AFST’s behavior with their peers less often than they used to. In addition, workers believed that they are now less likely to disagree with the AFST than they once were, having acclimated to using the tool and \textbf{having given up on the idea that their feedback can contribute meaningfully to improving the tool’s usefulness}. 
}

\edit{
So far, we have described how workers’ trust and reliance upon the AFST is influenced by organizational decisions relating to model transparency, choices of performance measures, and processes for overriding or providing feedback on algorithmic recommendations. In the next section, we describe how decisions about the design of the AFST model itself influenced workers’ perceptions and practices with the tool.
}

\subsection{\edit{Navigating value misalignments between algorithmic predictions and human decisions}} \label{valueMisalign}
\edit{
Workers often described differences between the criteria that they personally use to make decisions (ensuring that children are safe in the near term) versus the targets that the AFST predicts (risk of particular adverse outcomes over a two-year timespan). Due to these misalignments, several workers did not view the current version of the AFST as particularly relevant to the decisions they need to make day-to-day.
}

\subsubsection{\edit{Workers’ awareness of misalignments between the AFST’s predictive targets and their own decision targets}} 

Workers' focus on immediate safety and short-term risk has long been a point of contention with administration at the Allegheny DHS, and it has often been attributed to a lack of trust in the AFST. For example, in a 2018 interview, a DHS deputy director expressed that screeners \textit{``want to focus on the immediate allegation, not the child’s future risk a year or two down the line [...] Getting them to trust that a score on a computer screen is telling them something real is a process’’} \cite{hurley2018can}. \edit{While this narrative aligns with some of our findings, we found that \textbf{workers’ concerns around these misalignments extend well beyond a lack of trust in the underlying technology.} Some workers disagreed with the very idea of making screening decisions based on predictions of longer-term risk, perceiving this problem formulation as \textbf{fundamentally misaligned with their roles and responsibilities.} Other workers viewed the AFST’s focus on longer-term outcomes as \textbf{complementary to their own focus on immediate safety concerns.} However, while these workers were open, in theory, to the idea of factoring in longer-term risk into their decision-making, \textbf{they felt confused about how exactly they were expected to do so in practice.} 
}

\edit{
Both supervisors and call screeners described that they make decisions based on evidence of immediate risks to a child’s safety, in contrast to the algorithm’s focus on predicting longer-term risk. One supervisor, S2, worried that call screeners may interpret the AFST score as predicting ``risk'' without considering that its definition of risk does not necessarily imply immediate safety risk to a child. However, call screeners exhibited some awareness of the AFST’s notion of risk. For example, one call screener noted that:
\begin{quote}
\textit{``There are times where [knowing] the risk of removal within two years is not feeling like it's super relevant to the decision that is needed. And it [has] very little to do with immediate safety or anything like that.''}
\end{quote}
}

\edit{
\textbf{Most workers expressed that the AFST plays a relatively minor, non-driving role in their decision-making processes, overall.} When explaining why they avoid relying on the algorithm, these workers invoked not only the AFST’s perceived limitations (described in Section ~\ref{beliefs}), but also concerns about the outcomes that the tool predicts. For example, C6 shared that they sometimes feel that their personal goals of keeping children safe clash with the algorithm’s prediction targets: \textit{``it just has always felt like the risk of removal in two years is inherently going to be increased by our involvement, because we're the only ones that can remove the children.’’} They also worried about possible contradictions between the advertised goals of the AFST versus its use of proxy targets such as re-referral and placement: }
\begin{quote}
\textit{``To me, the tool has always been described as something to identify the families that might typically slip through the cracks or that wouldn't be most obvious upon initial assessment [...] Like, those families [that are frequently investigated because of the AFST every time they are referred] are not [the] ones that are slipping through the cracks and they're tripping on every single crack that they seem to encounter.’’} (C6)
\end{quote}

\edit{\textbf{Some workers went so far as to claim that they \textit{never} change their decisions based on the AFST score.}} For example, C4 said, \textit{``I look at the score. I often, you know, am in agreement with it. I think it, you know, does a good job trying to pull everything together and come up with the best possible solution with the score. [...] No [the AFST score does not change my recommendation]. I decide whatever I think.’’} C2 simply said, \textit{``I hate it [...] I don't think it should have a role, period, honestly''}. Others, like C6, described that when making recommendations, \textit{``[The AFST] doesn't feel like much more than a nudge in one direction or the other.’’} With that said, \textbf{observations during our contextual inquiries indicate that some of these participants may be underestimating the influence the AFST has on their decisions}. For example, C1 had claimed that \textit{``I personally, haven't had an instance where it caused me to change my recommendation.’’} \edit{However, in alignment with findings discussed in Section ~\ref{orgImpacts} regarding the impacts of perceived organizational pressures on workers’ practices, during a contextual inquiry, the worker was observed changing their screening recommendation from screen-out to screen-in after seeing the AFST score on a referral, explaining \textit{``I have to recommend it’’} because the AFST score was high (even though it was not high enough to require a mandatory screen-in). }

\subsubsection{\edit{Worker beliefs about complementarity between the AFST and themselves}} \label{beliefCompl}
\edit{Despite their awareness of misalignments between what the AFST predicts and the actual decisions they are trying to make, some workers saw value in using the AFST. Although workers were aware of potential biases and limitations of the AFST, \textbf{some believed that using the AFST might help to mitigate some of their personal biases.}} For example, C6 believed that the AFST score may help them mitigate their personal biases arising from cultural differences between themselves and the family:
\begin{quote}
\textit{``I think [...] a dirty house or something like that, I feel like those are often the ones where you're not sure if it is just a value or moral judgment and so, you know, the tool can probably weigh a little bit more in those circumstances [...] for those kinds of situations where, like, if I know that my personal bias may be underplaying a risk or something, [the AFST score] can be helpful.’’} (C6)
\end{quote}
Similarly, C9 reflected that \textit{``prior to the algorithm, things were screened out or not screened out [by the supervisor] and I would be surprised by that decision. I think, like, we're being a little more consistent with our decisions [now].’’}

\edit{Despite several workers describing clear value misalignments with the algorithm’s prediction targets, all supervisors and some call screeners (C5, C6, C7, C9) said that they \textbf{use the AFST score to help them make screening decisions in cases where they personally feel uncertain about a decision} or, for one participant, \textbf{when they are pressed for time}.} For example, S3 described that, for cases where they are initially uncertain about what screening decision to make, seeing the AFST score can help them more confidently make a decision: \textit{``Maybe it's something iffy, and then I run the score, and then the score is high. And then it's like, `Oh, yeah, definitely it should be assigned.’’’} Another supervisor, S2, agreed that the AFST score is helpful when the cases are not \textit{``straightforward,’’} providing physical discipline of a child as an example. Similarly, C6 said that they find the AFST score most helpful when they are \textit{``on the fence’’} about a decision. C9 noted that they usually follow the AFST’s recommendation more \textit{``if you're in a hurry.’’} 

That said, workers expressed that they have a harder time interpreting and using the AFST score when the score falls in the middle range (e.g., between 10 and 14), \edit{perceiving these cases as missed opportunities for the AFST to effectively complement their own judgment.} C5 explained that \textit{``that's where [supervisors] have the most problem, too [...they’ll] say every once in a while, like, `You got me [going] crazy over these f***ing yellow reports!’’’} 
\edit{
Given minimal transparency into what factors contribute to a mid-range score, and no information from the AFST beyond the score itself (Section ~\ref{beliefFormation}), workers generally disregarded the AFST’s outputs in these mid-range cases. \textbf{Workers expressed a desire to have the AFST communicate back, not \textit{``just a number’’} but also additional context explaining the AFST score}, for example, to assist them in interpreting middle range scores and in integrating the score into their decision-making.
}
\section{Discussion} \label{Discussion}
\edit{
As child welfare agencies increasingly adopt ADS to assist social workers’ day-to-day work \cite{aclu2021family, brown2019toward, Chouldechova2018, nissen2020social, saxena2020human}, it is critical that we understand workers’ experiences with these systems in practice. In this paper, we conducted the first in-depth qualitative investigation in the literature of child welfare workers' current practices and challenges in working with a prominent ADS (the AFST) day-to-day. The AFST context has been frequently studied in recent years (e.g., \cite{cheng2021soliciting, Chouldechova2018, De-Arteaga2020, vaithianathan2021using}), and public sector agencies are beginning to look to the AFST as an example of what AI-assisted decision-making can or should look like in child welfare and similar contexts \cite{aclu2021family}. However, most prior research on the AFST has relied on retrospective quantitative analyses of workers’ decisions, without an understanding of how workers actually integrate the AFST into their decision-making on-the-ground.}

\edit{
Through a series of contextual inquiries and semi-structured interviews, we observed ways in which workers’ reliance and trust on the AFST are guided by (1) their knowledge of rich, contextual information beyond what the underlying AI model captures, (2) their beliefs about the AFST’s capabilities and limitations in relation to their own, (3) organizational pressures and incentives that they perceive around the use of the ADS, and (4) their awareness of misalignments between the ADS's predictive targets versus their own decision-making objectives.
We found that, although workers at this agency had been using the ADS continuously for nearly half a decade, the system remained a source of tension for many workers, who perceived the system's current design as a missed opportunity to effectively complement their own abilities. Overall, our findings complicate narratives from prior literature about how the AFST and similar ADS tools may fit (or fail to fit) into workers’ day-to-day decision-making. These findings add to ongoing discussions in the literature, pointing to the need for a broader re-consideration of how ADS should be designed, evaluated, and integrated into future public sector contexts \cite{ammitzboll2021street, HoltenMoller2020, Saxena2021}. }

\edit{
In this section, we summarize each of our main findings, discuss how they extend or contrast with prior literature, and provide implications for the study of human-AI decision-making in child welfare and beyond.
Recent work in the area of human-AI interaction has highlighted that in many real-world decision-making settings, humans and AI systems may have access to \textit{complementary} information, opening potential for each to help overcome the other's limitations and blindspots (e.g., \cite{alkhatib2019street,De-Arteaga2020,Holstein,kleinberg2018human}). We found that in the AFST context, to calibrate their reliance on the AFST score in particular instances, \textbf{workers rely heavily upon their own knowledge of qualitative, contextual details of a given referred case, which are not captured by the administrative data that the AFST model uses.} We observed that workers frequently attempted to compensate for gaps and limitations in algorithmic predictions by drawing upon rich, causal narratives they had formed about a referred case, using their unique understanding of case-specific context. These qualitative findings validate hypotheses posed in earlier work by De-Arteaga et al. (2020). Through retrospective quantitative analyses of AFST-assisted decisions, these authors found that workers were able to reliably detect and override erroneous AFST recommendations. To explain how workers were able to do this, the authors speculated that the workers may have been cross-checking algorithmic outputs against other relevant information that was available to them but not to the AFST model \cite{De-Arteaga2020}. By contrast, much prior research on human-AI decision-making has focused on settings---often in the context of online crowdsourcing or laboratory-based studies---where the AI system has access to a \textit{superset} of the task-relevant information that is available to the human \cite{bansal2021does,De-Arteaga2020,LurieMulligan2020, Tan}. Our findings suggest caution in generalizing findings from such studies, as this setup may artificially advantage AI models over human decision-makers. Future research should seek to further understand how human decision-makers use complementary, model-external knowledge to calibrate their reliance on AI recommendations, as well as how future ADS tools might be designed to support them in doing so effectively. In addition, as discussed below, more research is needed to understand how human experts can be supported in \textit{communicating} relevant contextual knowledge to an AI model, to inform algorithmic recommendations in a given case (cf. \cite{De-Arteaga2021})---a desire expressed by workers in our study.}

\edit{
Despite having used the AFST day-to-day for multiple years, we found that \textbf{most workers knew very little about how the AFST works, what data it relies on, or how to work with the tool effectively.} Workers had received minimal training on how to use the AFST, and had almost no formal insight into how the AFST worked (Section ~\ref{beliefFormation}). These findings align with recent discussions in the human-AI interaction literature, suggesting that ADS tools are often introduced into professional contexts without adequate onboarding and training for the human decision-makers who are asked to work with them day-to-day (e.g., \cite{Bansal2019,cai2019hello}). Allegheny County is outwardly transparent about the AFST’s design and development process, providing much public-facing documentation on the web (e.g., \cite{afstfaq, AFSTdocumentation}). Indeed, this level of openness has contributed to the amount of influence and attention the AFST has received from researchers in HCI and machine learning, the popular press, and other public sector agencies around the world. However, this external transparency is not mirrored internally: workers are intentionally given limited information about the AFST model to avoid facilitating `gaming the system’ behaviors, where workers strategically manipulate data inputs to the algorithm, in order to get a desired output. Although this is a real concern in algorithm-assisted child welfare decision-making contexts (e.g., \cite{Saxena2021}), we observed that \textbf{given minimal information about the AFST, workers improvised ways to learn more about the tool themselves, building up sophisticated yet imperfect intuitions about the AFST’s behavior.} Workers drew upon their beliefs about the AFST to calibrate their reliance on algorithmic recommendations in particular instances---including what predictive features the AFST model uses, what influence each feature has, the kinds of cases for which the model is likely to be more or less reliable, and so on. Given this influence of workers’ informal beliefs about the AFST upon their use of the tool, it is unclear whether the agency’s original goals behind limiting transparency into the model were actually achieved. Yet this lack of transparency had other consequences: workers tended to be distrustful of the tool overall, and they did not feel equipped to work with it effectively. }

\edit{
To learn about the AFST’s behavior in the absence of formal training, workers engaged in various forms of what Shen, DeVos et al. (2021) have called \textit{``everyday algorithm auditing’’}: the ways that users detect, understand, and interrogate machine behaviors via their day-to-day interactions with algorithmic systems \cite{shen2021everyday}. Whereas these authors focused on characterizing the ways that members of large online communities come together to collectively make sense of algorithmic behavior, our findings present a case study of workers engaging in similar behaviors within a more intimate organizational setting. While readers may come away from prior quantitative research on AFST-assisted decision-making with images of workers as lone decision-makers, working in isolated cubicles, we found that workers often discussed the AFST model’s behavior with each other, to collectively understand unexpected scores. Through this collective deliberation, \textbf{workers shape each others’ beliefs about the model over time and influence each others’ decisions for specific cases} (Section ~\ref{beliefFormation}). Workers discussed AFST scores amongst themselves frequently and informally, even though the official screening protocol assigns only one call screener and one supervisor to attend to a given case. Workers also engaged in collaborative ``guessing games,’’ honing their abilities to predict the AFST score by guessing what the AFST score would be in particular cases, and then viewing the actual score. To learn about the influences of particular features on the AFST score, workers also frequently tinkered with the tool, for example by adjusting the inputs to the AFST model, recomputing the AFST score, and examining how the score changed. Finally, we observed that workers drew upon their knowledge of context and qualitative details of a given referred case in order to learn about the AFST’s accuracy over time---though it is important to note that the notion of ``accuracy’’ experienced by workers was not always aligned with that of the model. We expect that workers in other public sector contexts may engage in similar strategies to build up collective understandings of an ADS’s capabilities, limitations, and overall functioning---perhaps especially in contexts where workers are \textit{expected} to make AI-assisted decisions collaboratively, but are lacking relevant information about the AI model (e.g., \cite{Saxena2021,zytek2021sibyl}). Prior research investigating ways to improve AI-assisted decision-making, both in the AFST context and beyond, has often focused on improving interactions between a single human and an AI system. Our findings highlight the importance of studying how \textit{collaborative} decision-making---particularly in real-world organizational settings, in the presence of existing interpersonal and power relations among decision-makers---impacts how people rely upon and make sense of AI models.} 

\edit{
Consistent with our findings that workers were generally skeptical and distrustful of the AFST, we observed that workers’ decisions to follow or contradict the AFST score were often guided by factors other than their ``trust’’ in the AFST model. Whereas prior work has often studied human reliance on ADS tools as a function of an individual’s trust ~\cite{bansal2021does, buccinca2021trust, lee2004trust, Poursabzi-Sangdeh2021}, this conceptualization overlooks the potential impacts that organizational pressures and incentive structures may have on workers’ reliance, independent of their trust in the technology itself. In the AFST context, \textbf{workers perceived organizational pressures, both from the administration and from caseworkers downstream of their decision-making, that influenced their reliance on the AFST} (Section ~\ref{orgImpacts}). On the one hand, workers described complying with the AFST’s recommendation, against their own best judgment, to avoid being accused by the administration of disagreeing with the AFST ``too often’’ or of ``not using the tool.’’ On the other hand, workers considered caseworker load when deciding whether to screen out some ambiguous cases. Future research should seek to better understand the impacts that organizations’ internal messaging, performance measures, and policies have on the ways workers rely upon ADS tools in practice. In Section ~\ref{DesignImplications}, we discuss potential model-level and organization-level design implications based on these findings. In addition, based on their experiences, \textbf{workers came to perceive that they had little agency to shape the way the AFST is used within their organization, or to improve the reliability and accuracy of the AFST model itself}. As a result, workers shared that they have critical discussions about the AFST’s behavior with their peers less often than they used to. Workers also believed that, overall, they are now less likely to disagree with the AFST than they once were, having acclimated to using the tool and having come to believe that their disagreement may be looked down upon by the administration. In alignment with recent calls for greater consideration of temporal factors in the study of human-AI decision-making (e.g., \cite{lai2021towards}), our findings suggest that future research should investigate how human reliance upon ADS tools may evolve over time, particularly in real-world organizational settings.} 

\edit{
Importantly, we observed that despite having minimal formal training around the AFST, workers \textit{were} aware of \textbf{misalignments between the AFST’s predictive targets and their own decision-making objectives}, and that workers \textbf{took these misalignments into account when making AFST-assisted decisions} (Section ~\ref{valueMisalign}). Whereas workers generally focused on ensuring children’s immediate safety and considering short-term risk when making their decisions, the AFST was intentionally designed to \textit{complement} workers’ near-term focus by predicting proxies for adverse outcomes over a much longer (two-year) timespan \cite{AFSTdocumentation,De-Arteaga2021,vaithianathan2021using}. The goal was to nudge workers to consider not only immediate risk and safety, but also a forecast of longer-term risk, calculated based on available administrative data. However, we found that \textbf{it was unclear to workers how exactly they were expected to take the AFST’s assessments of long-term risk into account in ways that complemented their own judgment.} }
\edit{
While recent research in HCI and machine learning has begun to explore how to support human-AI complementarity in decision-making---i.e., configurations of humans and AI systems that yield better decisions in combination than either could achieve alone \cite{bansal2021does, De-Arteaga2020, Holstein, Kamar2016, Tan}---our findings drive home that it is not enough to have model targets that complement workers’ decision-making objectives \textit{in theory}. Human decision-makers must also understand, at a detailed level, how an ADS tool can actually complement their abilities in practice. As Saxena et al. (2021) emphasize \cite{Saxena2021}, human decision-makers must also be ``bought in’’ to the idea that their use of a given ADS tool will improve decision-making in ways that are meaningful to them. In turn, this requires that, \textit{even if} an AI tool is intended to complement human abilities by pushing them to consider complementary goals or objectives (cf. \cite{holstein2020conceptual}), \textbf{there must still be value alignment between humans and the AI model at a high-level.} Yet in the AFST context, we observed fundamental human-AI value misalignments. While some workers were open to the idea that the AFST could complement their judgment and help to mitigate some of their personal biases, other workers disagreed with the very idea of making screening decisions based on predictions of longer-term risk, perceiving this problem formulation as fundamentally misaligned with their roles and responsibilities. These observations echo findings from Saxena et al. (2021), who observed that, in agencies using simpler forms of algorithmic decision support, workers sometimes perceived conflicts between algorithmic recommendations and the ways they are actually trained to do their jobs. In addition to disagreeing with the timescale on which the AFST makes its predictions, workers were also uncomfortable with the AFST’s use of proxy targets such as re-referral and placement, viewing the use of these proxies as misaligned with their personal values. }

\edit{
Building upon prior approaches aimed at understanding and addressing misalignments between different stakeholders’ values during the design phase for ADS tools (e.g., \cite{HoltenMoller2020,Smith2020}, future research should further explore \textbf{how diverse stakeholders (e.g., decision-makers, affected populations, data scientists, and domain experts) can be effectively engaged in shaping \textit{model-level} design choices for an ADS}, such as specific choices of predictive targets or performance measures \cite{holstein2020replay, subramonyam2021towards, zhu2018value}. Future research should also explore \textbf{how best to support complementary human-AI performance in practice, in real-world organizational contexts}. This could involve, for example, investigating what other prerequisites and enabling conditions exist for human-AI complementarity, considering factors not just at the level of an AI model or individual human behavior, but also factors at the levels of groups of people or organizations. Finally, our findings suggest that when studying predictive accuracy or effective decision-making in the AFST context and other public sector decision-making contexts, it is \textbf{critical to account for potential differences in predictive targets and decision objectives of human decision-makers versus AI systems}. For instance, the misalignments observed in the AFST context complicate any argument that either the AFST or human workers are more accurate. Without an understanding of workers’ objectives when making predictions and decisions, these kinds of comparisons risk evaluating workers’ performance on a task that they are not actually performing.}
\section{Design Implications} \label{DesignImplications}
\edit{
Based on our findings, we provide the following design implications, intended for public-sector agencies deploying or maintaining ADS tools and for researchers exploring ways to design more effective ADS. We note that the following design implications are relevant only when there is good reason to expect benefits of having an ADS in the first place, which outweigh potential harms \cite{pierre2021getting, green2021flaws, baumer2011implication}. }
\begin{itemize}
    \item \edit {Support workers in using their expertise to improve an ADS's performance, whether by providing direct feedback to the ADS (e.g., communicating relevant contextual knowledge to inform algorithmic recommendations, for a given case), or by allowing the ADS to learn from important patterns across multiple workers' override decisions over time, as explored in De-Arteaga et al. (2021) ~\cite{De-Arteaga2021}.}
    \item \edit{Design training tools and interfaces that support workers in understanding the \textit{boundaries} of an ADS's capabilities (e.g., where it is likely to err and what factors it \textit{is not} able to take into account) \cite{mozannar2021teaching,Bansal2019,cai2019hello}. For example, decision support interfaces might be explicitly designed to support workers in exploring possible reasons for discrepancies between an ADS's statistical reasoning versus their own clinical judgment in a particular case (e.g., addressing workers' desires to have the AFST communicate back additional context explaining the score, as described in \ref{beliefFormation} and \ref{beliefCompl}).}
    \item \edit{Support open cultures for critical discussion around AI-assisted decision-making amongst workers using ADS. Providing transparency on ADS goals and algorithmic processes, as discussed above, can enable workers to engage in better informed discussions, decreasing the risk that such conversations will simply propagate misleading folk theories about an ADS. Moreover, supporting collaborative sensemaking and decision-making can encourage workers to surface insights about potential ADS limitations and improvements. \item Provide workers with balanced and contextualized feedback on their decisions. Feedback should capture the collaborative decision \textit{process} between workers and the ADS alongside the decision \textit{outcomes}. Both forms of feedback can be extremely valuable for learning, yet are scarce in many AI-assisted public-sector decision-making contexts (including the AFST context). Any resulting measures should be communicated to workers with explanations that account for both strengths and shortcomings of workers’ judgements. Such feedback may be critical for workers to reflect upon and improve their judgment and algorithmic reliance over time. }
    \item \edit{Relatedly, find ways to co-design measures of decision quality with the workers themselves. By designing measures of decision quality that workers find personally meaningful and important, workers may be more motivated to actually use these measures to reflect upon and improve their own decision-making. In addition, workers may be more likely to notice potential shortcomings in measures of decision quality, and may be well-equipped to notice when a given measure risks creating counterproductive incentive structures (e.g., that influence their reliance practices, as discussed in Section \ref{OrgMeasures}). }
    \item \edit{Clearly communicate, and in points of disagreement, collaboratively determine with social workers how decision-making power should be distributed across workers and the ADS. Building an organizational culture in which human decision-makers feel that their expertise is relevant and valued may preface building effective human--AI partnerships. Therefore, decision-making roles between workers and ADS should be guided by an empirical understanding of workers’ unique strengths (and limitations) in complementing the ADS.}
    \item \edit{Explore methods to support diverse stakeholder involvement in shaping an ADS’s model-level design decisions. For example, find methods to co-design an ADS’s predictive targets with decision-makers, affected community members, data scientists, and other domain experts, while considering how to mitigate disagreements ~\cite{HoltenMoller2020, Smith2020}, especially when prominent power dynamics are at play. Diverse stakeholder deliberation and discussion on model-level design decisions may help mitigate value misalignments between models and relevant stakeholders before the ADS is deployed.}
    \item \edit{Factor in a set of work context-specific considerations when designing ADS predictive targets that aim to complement a worker’s decision targets and processes. For example, beyond considering whether a human’s and an AI’s limitations and strengths might complement one another in a vacuum, consider the organizational factors that may influence reliance decisions in practice.}
    \item \edit{When assessing the effectiveness of human-AI decision-making, be cautious not to adopt evaluation approaches that artificially disadvantage either the human or the AI. For example, carefully consider whether the metrics used to evaluate human-AI performance are evaluating the actual outcomes that the human versus the AI is optimizing towards, when making predictions or decisions.}

\end{itemize}
\section{Conclusion} \label{Conclusion}
Our findings highlight critical opportunities for future research, towards re-thinking and re-designing the \edit{interfaces, models, and organizational processes that shape the ways ADS tools are used in child welfare and other public sector decision-making contexts.} Future research across HCI, machine learning, and social work should explore design opportunities at all three of these levels, both separately and in combination. In addition, future research should seek to inform the design of human--AI partnerships in child welfare by understanding and reconciling \edit{design values and goals} across a broader ecosystem of relevant stakeholders, including social work administrators and agency leadership, caseworkers \cite{Saxena2021}, families and other affected community members \cite{brown2019toward,cheng2021soliciting}, and ADS developers \cite{HoltenMoller2020}. In particular, future research should explore how multiple stakeholders within this ecosystem might be meaningfully involved across various points of the design, development, deployment, use, and maintenance lifecycle for ADS, to resolve value misalignments and \edit{to better serve the needs of families and child welfare workers.}
\begin{acks}
\blue{We thank the call screeners and supervisors at Allegheny County for their time and valuable input that shaped this research. We also thank leadership at Allegheny County for their strong commitment to transparency, and opening their doors to allow external researchers to closely observe their practices. Finally, we thank our anonymous reviewers for their dedication and thorough feedback that substantially improved this paper. This work was supported by the National Science Foundation (NSF) under Award No. 1939606, 2001851, 2000782 and 1952085 and the Carnegie Mellon University Block Center for Technology and Society Award No. 53680.1.5007718.} %check if these are correct.
\end{acks}

\bibliographystyle{ACM-Reference-Format}
\bibliography{references}

\end{document}